\documentclass{emulateapj}

\newcommand{\degg}{\hbox{$^\circ$}}

\newcommand{\et}{et al.\ }

\newcommand{\ls}{\mathrel{\hbox{\rlap{\hbox{\lower4pt\hbox{$\sim$}}}\hbox{$<$}}}}
\newcommand{\gs}{\mathrel{\hbox{\rlap{\hbox{\lower4pt\hbox{$\sim$}}}\hbox{$>$}}}}
\newcommand{\grb}{GRB~031203}

\begin{document}

\title{The discovery of an evolving dust scattered X-ray halo 
around \grb}
\shorttitle{An X-ray halo around \grb}
\author{ S. Vaughan\altaffilmark{1}, 
R. Willingale\altaffilmark{1},
P. T. O'Brien\altaffilmark{1},
J. P. Osborne\altaffilmark{1}, 
J. N. Reeves\altaffilmark{2,3}, 
A. J. Levan\altaffilmark{1}, 
M. G. Watson\altaffilmark{1},
J. A. Tedds\altaffilmark{1}, 
D. Watson\altaffilmark{4},
M. Santos-Lle\'o\altaffilmark{5},
P. M. Rodr\'{\i}guez-Pascual\altaffilmark{5},
N. Schartel\altaffilmark{5}
} 
\email{sav2@star.le.ac.uk}

\altaffiltext{1}{Department of Physics \& Astronomy, University of
Leicester, Leicester LE1 7RH, United Kingdom.}

\altaffiltext{2}{Laboratory for High Energy Astrophysics, Code 662,
NASA Goddard Space Flight Center, Greenbelt Road, Greenbelt, MD 20771,
USA.}

\altaffiltext{3}{Universities Space Research Association}

\altaffiltext{4}{Niels Bohr Institute for Astronomy, Physics and
Geophysics, University of Copenhagen, Julian-Maries Vej 30, DK-2100,
Copenhagen \O, Denmark.}

\altaffiltext{5}{XMM-Newton SOC, Villafranca, Apartado 50727, E-28080
Madrid, Spain.}

\begin{abstract}
We report the first detection of a time-dependent, dust-scattered
X-ray halo around a gamma-ray burst. \grb\ was observed by {\it
XMM-Newton} starting six hours after the burst. The halo appeared as
concentric ring-like structures centered on the GRB location. The
radii of these structures increased with time as $t^{1/2}$, consistent
with small-angle X-ray scattering caused by a large column of dust
along the line of sight to a cosmologically distant GRB. The rings are
due to dust concentrated in two distinct slabs in the Galaxy located
at distances of $880$ and $1390$ pc, consistent with known Galactic
features. The halo brightness implies an initial soft X-ray pulse
consistent with the observed GRB. 

\end{abstract}


\section{Introduction}

It has long been realized that the small-angle scattering of X-rays by
dust grains can result in a detectable X-ray ``halo'' around a distant
X-ray source (Overbeck 1965), with a radial intensity distribution
that depends on the dust properties and location. This phenomenon was
first detected by Rolf (1983) using data from {\it Einstein} and
confirmed by later observations (e.g. Mauche \& Gorenstein 1986;
Predehl \et 1991).

In the case of Gamma-Ray Bursts (GRBs), the transient nature of the
burst combined with the large initial X-ray flux means X-ray
scattering may be visible for a short period of time with a relatively
high surface brightness. As the transient flux passes through a dust
slab between the observer and the GRB, the scattering process
introduces a time-delay in which the X-rays at larger angles to the
line of sight arrive increasingly delayed with respect to the
non-scattered X-rays. Thus, the dust allows us to view the GRB X-ray
flux at earlier times. The time-dependent X-ray halo around the GRB
can, in principle, be used to provide detailed information on the
location, spatial distribution and properties of the dust and
the distance to and brightness of the GRB (e.g. Tr\"umper \&
Sch\"ofelder 1973; Klose 1994; Miralda-Escud\'e 1999; Draine 2003).
Images of transient X-ray sources seen in scattered light are
analogous to ``light echos'' seen around some supernovae (e.g. SN
1987A, Xu, Crotts \& Kunkel 1995).

Scattered X-rays observed at an angle $\theta$ from the line-of-sight
to a GRB arrive after the direct emission with a time-delay $\tau_s$
given by:

\begin{equation}
\tau_s = \frac{(1+z_s) D_s D_g \theta^2 }{2 c D_{gs}},  
\end{equation}

where $z_s$ and $D_s$ are the redshift and angular diameter distance
of the scattering dust, $D_g$ is the angular diameter distance
of the GRB and $D_{gs}$ is the angular diameter distance from the dust
to the GRB (Miralda-Escud\'e 1999). 

Here we discuss the case of \grb\ observed by {\it XMM-Newton} on
two occasions shortly after the burst. The first observation revealed the
first dust-scattered X-ray halo detected around a GRB.

\section{Observations and data analysis}

\begin{figure*}[!t]
\centering
\epsscale{0.65}
\rotatebox{270}{
\plotone{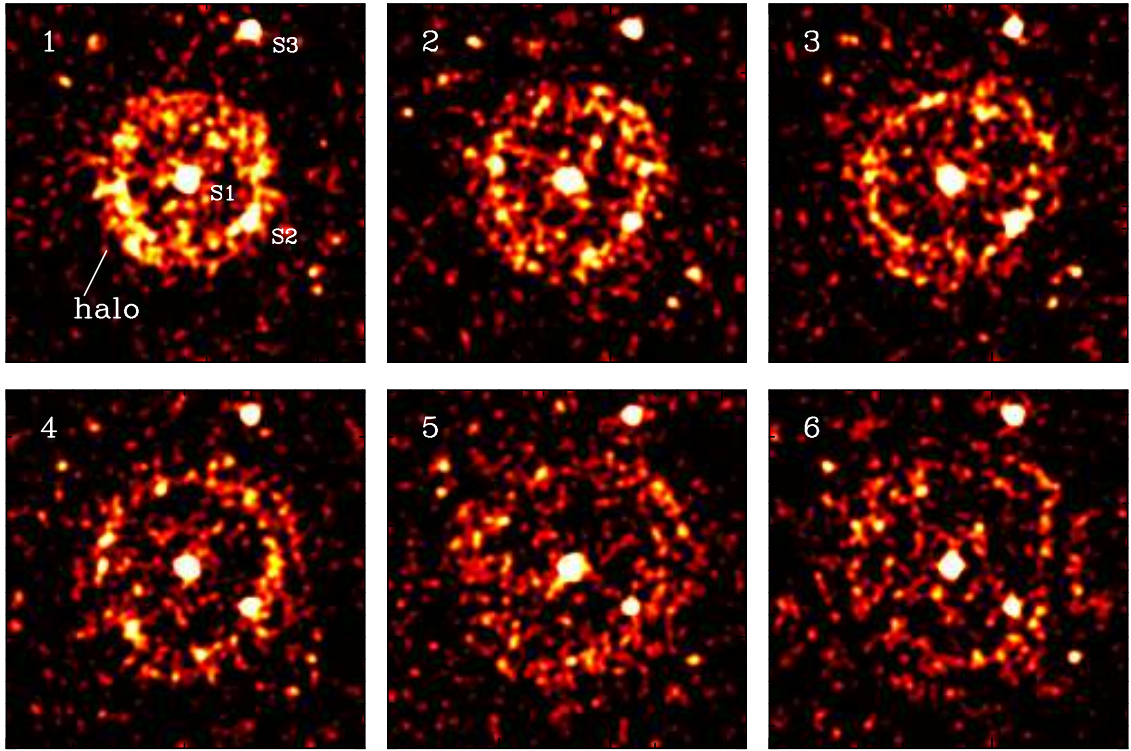}
}
\caption{
Combined MOS images of \grb\ covering the $0.7-2.5$~keV range,
spanning $10\arcmin$ on a side. The data were divided into
$10$ contiguous time intervals lasting $5780$~s. The images were
obtained from the first six time slices are shown (smoothed
using a $6$~arcsec Gaussian kernel). The three
brightest point sources (S1, S2, S3) are marked.}
\vspace{-0.5 cm}
\end{figure*}

\grb\ was detected by the IBIS instrument on {\it Integral} on
2003 December 3 at 22:01:28 UT as a single peaked burst with a
duration of 30 s and peak flux of $1.3 \times 10^{-7}$ erg s$^{-1}$
cm$^{-2}$ in the 20--200 keV band (Gotz \et 2003; Mereghetti \& Gotz
2003). A 58 ks {\it XMM-Newton} observation of the field began at 2003
December 4, 04:09:29 UT. Several X-ray sources were detected within
the $2.5\arcmin$ radius {\it Integral} error-circle (Santos-Lle\'o \&
Calder\'on 2003), the brightest of which (S1) appeared to fade through
the observation (Rodr\'{\i}guez-Pascual \et 2003) and was interpreted
as the X-ray afterglow of \grb. We refined the astrometric
position of the afterglow by performing a cross-correlation with the
USNO-A2 catalog, the improved S1 position was (J2000) RA=$08^{\rm h}
02^{\rm m} 30.19^{\rm s}$, Dec$=-39\degg$~$51$\arcmin~$04.0\arcsec$
($1\sigma$ error 0.7\arcsec, Tedds et al. 2003).

The position in the Galactic plane ($l=255\degg, b=-4.6\degg$) results
in high optical extinction ($E(B-V) =1.0$, Schlegel, Finkbeiner \&
Davis 1998). Optical/IR observations failed to locate an afterglow.
Radio observations did locate a transient source location consistent
with the X-ray afterglow (Soderberg, Kulkarni \& Frail 2003), and
optical imaging and spectroscopy revealed a candidate host galaxy at
$z=0.105$ which shows star-forming features typical of GRB hosts
(Prochaska \et 2003a,b). Our refined location for the X-ray afterglow
is within 0.5\arcsec\ of the optical galaxy and consistent with the
position of the radio source. 

During the first {\it XMM-Newton} observation the GRB was observed on
axis with all EPIC instruments in full-frame modes (Str\"{u}der \et
2001; Turner \et 2001). A second 54 ks {\it XMM-Newton} observation
was obtained on 2003 December 6 in which the GRB was $\approx
6\arcmin$ off axis. The observational details and the GRB spectrum and
light-curve are discussed in detail in Watson \et (2004).

The extraction of science products followed standard procedures using
the {\it XMM-Newton} Science Analysis System ({\tt SAS v5.4.1}).  The
EPIC data were processed using the {\tt SAS} chains to produce
calibrated event lists and remove background flares. GRB afterglow
source counts were extracted from circular regions (radius
$34\arcsec$) and background counts were estimated from a large
off-axis region free from obvious point sources and the halo. The GRB
afterglow fades as $(t-t_0)^{-0.4}$, where $t_0$ is the time of the
GRB, which is unusually slow for GRB
X-ray afterglows.

\section{The X-ray Halo}

\begin{figure*}[!t]
\centering
\epsscale{0.65}
\rotatebox{270}{
\plotone{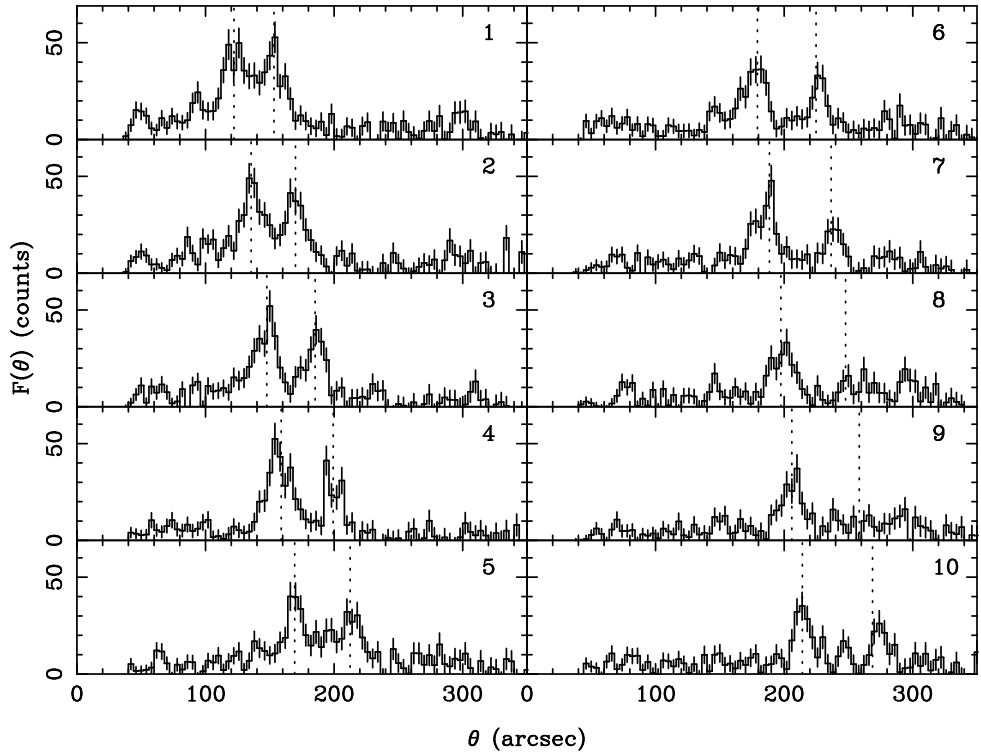}
}
\caption{
Radial profile of counts about the \grb\ afterglow from ten
contiguous time intervals (derived from the combined MOS
$0.7-2.5$~keV images). The afterglow itself (S1) and the
next two brightest points sources (S2, S3) were removed
prior to calculating the profiles (the positions of S2 and
S3 are marked in the upper panels).  The two peaks
corresponding to the two expanding rings are obvious. The
dotted lines denote the best-fitting $\theta \propto
(t-t_0)^{1/2}$ functions (see Fig.~3).}
\vspace{-0.5 cm}
\end{figure*}

The X-ray image from the first {\it XMM-Newton} observation reveals an
extended, circular halo concentric with the \grb\ afterglow (see
Fig.~1). This halo was seen in all three cameras of the EPIC
instrument, is unique to this observation, and is not due to scattered
optical or X-ray light within the instrument. Summing the entire
observation the halo had the form of a virtually complete ring
(Vaughan \et 2003), but splitting the observation into contiguous time
intervals revealed distinct rings which increased in radius
through the observation.

In order to better quantify this expansion, EPIC MOS images were
produced for ten contiguous time intervals of $5780$~s duration,
spanning the entire first observation, and binned to $4\arcsec$
pixels. We concentrate on the MOS data as the halo lies over several
chip gaps in the EPIC pn images, but the same halo was seen in both.
The radial profiles of the images about the afterglow were calculated
removing the three brightest point sources within the region of
interest (S1, S2 and S3 in Fig.~1) by ignoring all counts within
$40\arcsec$ from the centroid of each source. The radial profiles were
then background subtracted using the mean level from the
$350-400\arcsec$ annulus, which is well outside the detected halo. The
resulting count profiles as a function of radius are shown in Fig.~2.
The strongest peak moved outward from $\sim 120\arcsec$ to $\sim
220\arcsec$ during the observation, while a second peak moved from
$\sim 160$ to $\sim 270\arcsec$.

The radial positions of the two rings were measured from the local
maxima of the two peaks in the radial distributions. Fig.~3 shows the
change in radii as a function of time, and both were well fit by a
simple power-law ($\theta \propto (t-t_0)^{\alpha}$). Fixing $t_0 =
0$, the indices of the power-law expansions were measured to be
$\alpha=0.53\pm0.04$ and $0.51\pm0.05$ for the inner and outer rings
respectively (errors are 90\% confidence intervals), consistent with
the $\theta \propto (t-t_0)^{1/2}$ expansion predicted for a scattered
halo. As a consistency check we allowed $t_0$ to be a free parameter
but fixed $\alpha=0.5$ and refitted to derive limits of $t_0 =
2794_{-3178}^{+2765}$~s and $2005_{-2867}^{+2512}$~s for the inner and
outer ring respectively, both consistent with $t_0 = 0$. The second
{\it XMM-Newton} observation obtained 3 days after the burst showed no
evidence for any ringed structure. 

The distance to the GRB is large (adopting the redshift of the
putative host galaxy), the time-delay is short and $\theta$ is
relatively large, thus to a good approximation $D_g = D_{gs}$ and
$(1+z_s) = 1$ in equation (1), hence $D_s = 2c \tau_s / \theta^2$.
Assuming $\theta \propto (t-t_0)^{1/2}$ and $t_0 = 0$, the data in
Fig.~3 imply that the scattering dust slabs are located at distances
from the observer of $D_1 = 1388\pm32$ pc and $D_2 = 882\pm20$ pc,
corresponding to the inner and outer ring respectively. At these
distances, the ring diameters probe size scales of 2--3 pc in the
dust scattering medium.

The halo spectrum was extracted from an annulus with radii
$110-220\arcsec$ using only the first half of the observation; the
decreased surface brightness of the halo at later times makes spectral
extraction unreliable. A background spectrum was extracted from a
large off-axis region (avoiding the halo and other point sources).
Fig.~4 shows the halo spectrum compared to the afterglow spectrum. The
spectrum of the halo is much steeper than that of the afterglow.
Fitting with a simple absorbed power-law yielded photon indices of
$\Gamma = 1.98\pm0.05$ and $\Gamma = 3.03\pm0.14$ for the afterglow
and halo, respectively. The best-fitting absorbing column was $N_{\rm
H} = 8.8 \pm 0.5 \times 10^{21}$~cm$^{-2}$ when fixed to be the same
for both afterglow and halo spectra. This is larger than the Galactic
value of $N_{\rm H} = 6.1 \times 10^{21}$~cm$^{-2}$ (Dickey \& Lockman
1990). The specific fluxes at $t=36650$~s after the burst at $1$~keV
are $1.5_{-0.1}^{+0.2} \times 10^{-4}$ photons cm$^{-2}$ s$^{-1}$
keV$^{-1}$ for the afterglow and $6.0 \pm 0.4
\times 10^{-4}$ photons cm$^{-2}$ s$^{-1}$ keV$^{-1}$ for the halo.

To model the dust scattering medium, we have simultaneously fitted the
ten radial profiles shown in Fig.~2 using the differential and total
scattering cross sections of the Rayleigh-Gans approximation (Mauche
and Gorenstein 1986). This is a valid approximation if the grain
radius in $\mu$m is much less than the X-ray photon energy in keV
(Smith \& Dwek 1998). We assume that the dust is confined to two slabs
at distances $D_{1}$ and $D_{2}$ with thickness $\Delta D$. The halo
seen is created by the combination of the soft X-ray pulse from the
GRB and the subsequent afterglow. The observed angular width of the
rings results from a combination of the duration of the soft X-ray
pulse, $\Delta t$, $\Delta D$ and the point spread function (PSF) of
the telescope. The width of the PSF corresponds to $\Delta
t\sim 1000$ s or $\Delta D\sim100$ pc. Given that the GRB is very
short, $\sim30$ s (Mereghetti \& Gotz 2003), we assume that $\Delta t
\ll 1000$ s and hence, allowing for the PSF,
the observed broadening of the rings, $20\arcsec$, gives us an
estimate of the slab thickness, $\Delta D=130\pm50$ pc.

The angular and temporal distribution was well fitted using grains
with radii in the range $0.15<a<0.25$ $\mu$m. Smaller grains scatter to
much larger angles (not visible because the time elapsed since the
burst is too small) and the upper limit to the grain size distribution
is expected to be $\approx 0.3$ $\mu$m (Predehl et al. 1991). Using
this range of grain sizes, we require the initial GRB pulse and
afterglow to have a photon index of $\Gamma \sim 2.0$ (as observed for
the afterglow from 6 h) to match the observed photon index of the halo
($\Gamma = 3.0$).

\begin{figure}
\centering
\epsscale{0.75}
\rotatebox{270}{
\plotone{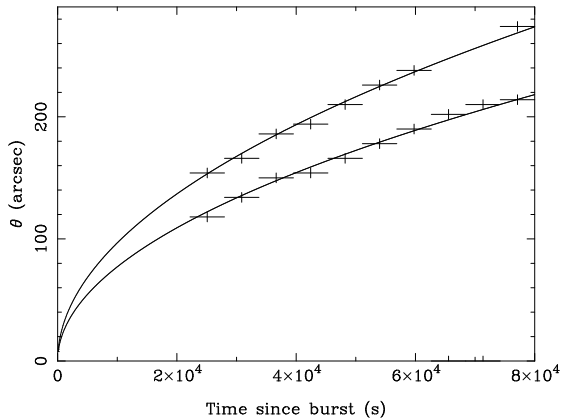}
}
\caption{
Expansion of the two rings around \grb\ with time.  The radius of each
ring was measured from the local maxima of the radial
profiles (Fig.~2). In both cases the expansion was
well-fitted with a functional form $\theta \propto
(t-t_0)^{1/2}$. The errors were assumed to be the width of
the bin size used for the radial profile measurement.}
\end{figure}

\begin{figure}
\centering
\epsscale{0.75}
\rotatebox{270}{
\plotone{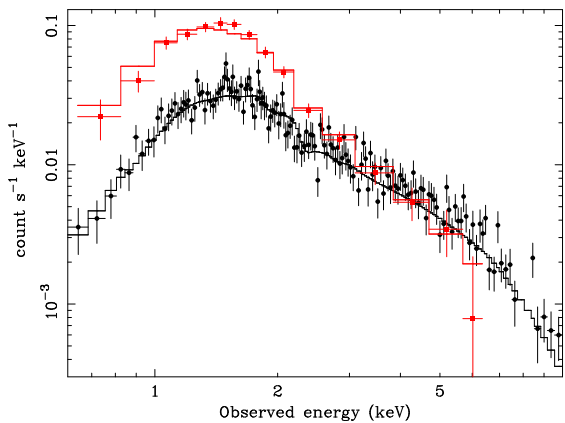}
}
\caption{
Absorbed power-law model fits to the EPIC pn spectra of the \grb\
afterglow (circles, thick line) integrated over the entire observation
and the dust halo (squares, thin line) integrated over the first half
($27895$~s) of the observation (during which the surface brightness
was high enough for a reasonable spectral extraction).  This clearly
demonstrates the spectrum of the halo is significantly steeper than
the afterglow.}
\end{figure}

The brightness of the halo depends on the product of the differential
scattering cross sections of the grains, the column densities of the
grains in the slabs and the integrated flux of the X-ray pulse and
afterglow. Interstellar extinction maps show structure in the GRB
direction, with a considerable increase to $A_V \approx 2$ mag at a
distance of $\sim 1.3$ kpc (Neckel \& Klare 1980).  This distance is
consistent with the location of the more distant dust slab we detect.
Adopting a mean grain size of $a=0.2$ $\mu$m, an $A_V = 2$ mag
corresponds to a grain column density of $1.5\times10^{8}$ cm$^{-2}$
(Mauche and Gorenstein 1986). The best fit to the ring structure
($\chi^{2}$/d.o.f. $=836/489$) results from dividing this column
density between the two slabs, $N_{1}\sim 1\times10^{8}$ and
$N_{2}\sim 0.5\times10^{8}$ cm$^{-2}$ (i.e. most of the dust is in the
more distant slab). Using these values and $a=0.20\pm0.05$, the
time-integrated flux required for the soft X-ray pulse is $1600\pm800$
photon cm$^{-2}$ keV$^{-1}$ at 1~keV.  Assuming a 30 s rectangular
pulse this is equivalent to $\sim 5$ Crabs; using $\Gamma = 2$ the
predicted GRB flux in the $20-200$~keV band is $2.4\pm 1.2$ photon
cm$^{-2}$ s$^{-1}$.  Given the assumptions (e.g. spectral shape) this
is in reasonable agreement with the value reported for \grb\ (peak of $\sim
1.2$ photon cm$^{-2}$ s$^{-1}$, Gotz et al. 2003; Mereghetti \& Gotz
2003). The X-ray burst is very bright compared to the afterglow (see
Watson \et 2004). Extrapolating the
decay of the afterglow (section~2) to the time of the GRB
accounts for only $\sim 2$ per cent of the halo emission.

The line of sight to \grb\ passes close to the center of the Gum
Nebula which appears as a $28\degg$ diameter sphere in H$\alpha$ centered at
$l=258\degg , b=-2\degg$ (Chanot \& Sivan 1983). The Gum Nebula is
likely to be a supershell created by repeated SNe explosions within
it. Reynoso \& Dubner (1997) detected a neutral gas disk
(H~{\textsc{i}}) associated with the Gum Nebula, located $\sim 500\pm
100$ pc from us, with a radius of $\sim 150$ pc.
The dust slab nearest to us may be associated with the rear face of 
the Gum Nebula which is known to have a concentration of molecular
clouds (Woermann, Gaylard \& Otrupcek 2001).

\acknowledgements

This paper is based on observations obtained with XMM-Newton, an ESA
science mission with instruments and contributions directly funded by
ESA Member States and the USA (NASA). SV and AJL acknowledge support
from PPARC, UK. We acknowledge the benefits of collaboration within
the Research and Training Network ``Gamma-Ray Bursts: An Enigma and a
Tool'', funded by the European Union under contract
HPRN-CT-2002-00294. We thank an anonymous referee for a constructive
report.

\end{document}